\documentclass[aps,prl,twocolumn,superscriptaddress]{revtex4-2}
\usepackage{amsmath,amssymb,amsbsy,graphicx,xcolor,times,subfigure,marvosym,color,bm,multirow,epstopdf}
\usepackage[latin1]{inputenc}
\begin{document}

\title{Low-energy edge signatures of the Kitaev spin liquid}

\author{Shang-Shun Zhang}
\affiliation{Department of Physics and Astronomy, The University of Tennessee, Knoxville, Tennessee 37996, USA}

\author{Cristian D. Batista}
\affiliation{Department of Physics and Astronomy, The University of Tennessee, Knoxville, Tennessee 37996, USA}
\affiliation{Neutron Scattering Division and Shull-Wollan Center, Oak Ridge National Laboratory, Oak Ridge, Tennessee 37831, USA}

\author{G\'abor B. Hal\'asz}
\thanks{This manuscript has been authored by UT-Battelle, LLC, under contract DE-AC05-00OR22725 with the US Department of Energy (DOE). The publisher acknowledges the US government license to provide public access under the DOE Public Access Plan (http://energy.gov/downloads/doe-public-access-plan).}
\affiliation{Materials Science and Technology Division, Oak Ridge National Laboratory, Oak Ridge, Tennessee 37831, USA}
\affiliation{Quantum Science Center, Oak Ridge, Tennessee 37831, USA}


\begin{abstract}

Recent experimental work indicates that the Kitaev spin liquid may be realizable close to zero magnetic field in exfoliated $\alpha$-RuCl$_3$ flakes, thus providing a more versatile setting for studying non-Abelian anyons. Here, we propose a robust nanoscale signature of the Kitaev spin liquid that results from its edge states and manifests in the low-energy spin dynamics. In particular, we highlight a singular peak in the dynamical spin structure factor of a zigzag edge whose energy scales linearly with only one component of the magnetic field. This sharp feature in the local spin dynamics directly reflects the bond-directional Kitaev spin interactions and, more generally, the projective symmetries of the Kitaev spin liquid. We demonstrate that our proposed signature survives in the presence of edge disorder as well as non-Kitaev interactions, and provide detailed guidelines for experimentally observing it using inelastic electron tunneling spectroscopy and color-center relaxometry.

\end{abstract}


\maketitle


\emph{Introduction.}---The Kitaev spin liquid~\cite{Kitaev-2006} is a highly entangled quantum phase of matter hosting two kinds of fractional excitations: gapless Majorana fermions (spinons) and gapped $\mathbb{Z}_2$ gauge fluxes (visons). Remarkably, this gapless spin liquid also has an exactly solvable representation in the ground state of the famous Kitaev honeycomb model~\cite{Kitaev-2006}. Moreover, in the presence of a magnetic field, the gapless Dirac points of the Majorana fermions are gapped out, while the $\mathbb{Z}_2$ gauge fluxes acquire non-Abelian quasiparticle statistics. Therefore, the resulting field-induced gapped phase---the non-Abelian Kitaev spin liquid---is important from the perspective of topological quantum computation~\cite{Kitaev-2003,Nayak-2008}.

The most promising candidate material for realizing the Kitaev spin liquid is arguably $\alpha$-RuCl$_3$~\cite{Plumb-2014,Kubota-2015,Sandilands-2015,Sears-2015,Majumder-2015,Johnson-2015,Sandilands-2016,Banerjee-2016,Banerjee-2017,Do-2017}. Indeed, it is hypothesized that bulk samples of this spin-orbit-coupled honeycomb magnet may support the non-Abelian Kitaev spin liquid for in-plane magnetic fields of $\sim 10$ T~\cite{Banerjee-2018,Kasahara-2018,Balz-2019,Yokoi-2021,Bruin-2022,Imamura-2024}, although the experimental evidence is highly controversial~\cite{Yamashita-2020,Czajka-2021,Lefrancois-2022,Czajka-2023}. At the same time, a recent work demonstrates that the field-induced behavior of monolayer $\alpha$-RuCl$_3$ is strikingly different from its bulk counterpart, and argues that the bond-directional spin interactions of the Kitaev honeycomb model are significantly enhanced in the monolayer form~\cite{Yang-2023}. Given the versatility of similar two-dimensional materials~\cite{Chaves-2020}, monolayer $\alpha$-RuCl$_3$ is then a viable platform for realizing the parent phase of the non-Abelian Kitaev spin liquid---the original gapless Kitaev spin liquid---already at zero magnetic field, thus providing a potential route to non-Abelian quasiparticle statistics at magnetic fields $\ll 10$ T. However, the experimental detection of the Kitaev spin liquid is especially difficult in a monolayer sample because traditional bulk techniques like inelastic neutron scattering are not available.

In this Letter, we propose a robust nanoscale signature for the Kitaev spin liquid that manifests in the low-energy edge dynamics and is revealed by a small magnetic-field perturbation. Specifically, due to its particular interplay with the honeycomb symmetries, the zigzag edge of the Kitaev spin liquid supports divergent spin fluctuations at an energy that scales linearly with only a single component of the applied magnetic field. This sharp feature in the edge dynamics directly reflects both the bond-directional spin interactions of the Kitaev honeycomb model and, more generally, the projective (i.e., fractional) symmetries of the Kitaev spin liquid~\cite{You-2012,Song-2016}. Indeed, while our main calculation is based on an ideal zigzag edge of the pure Kitaev model, we also demonstrate that its results are robust against both moderate edge disorder and non-Kitaev interactions; additional features produced by stronger edge disorder are discussed in a companion paper~\cite{Zhang-2024}. Experimental approaches for observing our proposed edge signature include inelastic electron tunneling spectroscopy~\cite{Yang-2023,Feldmeier-2020,Konig-2020,Bauer-2023,Takahashi-2023,Kao-2024a,Kao-2024b} as well as color-center relaxometry~\cite{Rondin-2014,Casola-2018,Chatterjee-2019,Gottscholl-2021,Liu-2022}.

\begin{figure}[b]
\includegraphics[width=0.9\columnwidth]{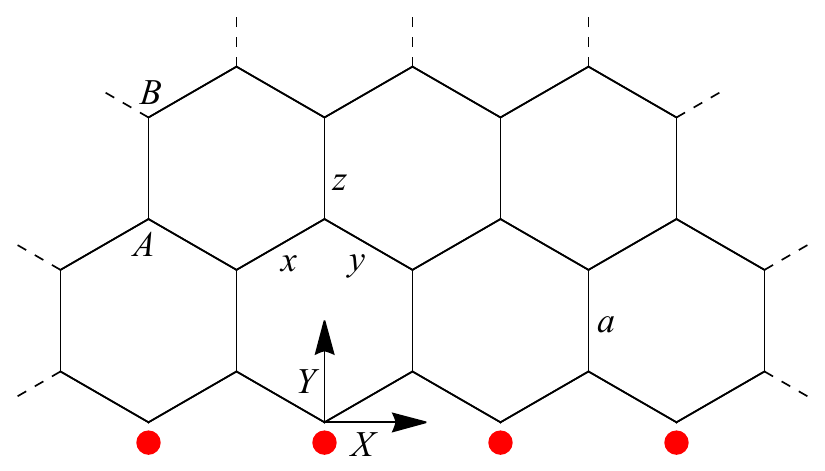}
\caption{Ideal zigzag edge of the honeycomb lattice. Dangling bond fermions $\tilde{b}_{\mathbf{r} \in \mathbb{D}_z}^z$ are marked by red dots. The honeycomb sublattices ($A,B$), the bond types ($x,y,z$), the bond length ($a$), and the dimensionless coordinates ($X,Y$) are also shown.} \label{fig-1}
\end{figure}

\emph{Model.}---We represent the Kitaev spin liquid with the exactly solvable Kitaev honeycomb model~\cite{Kitaev-2006},
\begin{equation}
\mathcal{H}_K = -K \sum_{\langle \mathbf{r}, \mathbf{r}' \rangle_x} \sigma_{\mathbf{r}}^x \sigma_{\mathbf{r}'}^x - K \sum_{\langle \mathbf{r}, \mathbf{r}' \rangle_y} \sigma_{\mathbf{r}}^y \sigma_{\mathbf{r}'}^y - K \sum_{\langle \mathbf{r}, \mathbf{r}' \rangle_z} \sigma_{\mathbf{r}}^z \sigma_{\mathbf{r}'}^z, \label{eq-H-1}
\end{equation}
and concentrate on an ideal zigzag edge of the underlying honeycomb lattice (see Fig.~\ref{fig-1}). We assume without loss of generality that the zigzag edge is perpendicular to the $z$ bonds. By decomposing the spins into Majorana fermions in the standard way, $\sigma_{\mathbf{r}}^{\alpha} = i b_{\mathbf{r}}^{\alpha} c_{\mathbf{r}}^{\phantom{\dag}}$ (with $\alpha = x,y,z$), the bond fermions $b_{\mathbf{r}}^{\alpha}$ are paired up into bond variables $u_{\mathbf{r}, \mathbf{r}'}^{\alpha} \equiv i b_{\mathbf{r}}^{\alpha} b_{\mathbf{r}'}^{\alpha} = +1$ that possess a finite energy gap $\Delta$ and represent the ground-state flux sector of the Kitaev model~\cite{Kitaev-2006}. The low-energy manifold then consists of the matter fermions $c_{\mathbf{r}}$ and the unpaired bond fermions $\tilde{b}_{\mathbf{r}}^z$ that correspond to ``dangling'' edge sites $\mathbf{r} \in \mathbb{D}_z$ with no $z$ bonds connected to them (see Fig.~\ref{fig-1}). The effective Hamiltonian in terms of these low-energy fermions is a simple first-neighbor hopping problem for the matter fermions,
\begin{equation}
H_K = K \sum_{\langle \mathbf{r} \in A, \mathbf{r}' \in B \rangle} i c_{\mathbf{r}} c_{\mathbf{r}'}, \label{eq-H-2}
\end{equation}
where $A$ and $B$ are the two honeycomb sublattices. Note that the dangling bond fermions $\tilde{b}_{\mathbf{r}}^z$ do not appear in this Hamiltonian as they are completely decoupled.

We also consider a small magnetic field described by a Zeeman term, $\mathcal{H}_h = -\vec{h} \cdot \sum_{\mathbf{r}} \vec{\sigma}_{\mathbf{r}}$, as a perturbation to reveal the low-energy edge dynamics of the Kitaev spin liquid. Assuming $|\vec{h}| \ll \Delta$, the leading-order effect of this perturbation is obtained by simply projecting it onto the low-energy manifold that corresponds to the ground-state flux sector. For a generic site $\mathbf{r} \notin \mathbb{D}_z$, each spin component $\sigma_{\mathbf{r}}^{\alpha}$ anticommutes with the corresponding bond variable, $u_{\mathbf{r}, \mathbf{r}'}^{\alpha}$, implying that its projection to the low-energy manifold vanishes. For a dangling site $\mathbf{r} \in \mathbb{D}_z$, however, the $z$ component $\sigma_{\mathbf{r}}^z$ is not associated with any bond variable and has a nontrivial projection that directly couples the dangling bond fermion $\tilde{b}_{\mathbf{r}}^z$ to the matter fermion $c_{\mathbf{r}}$~\cite{Kao-2021,Takahashi-2023,Kao-2024a,Kao-2024b}. Hence, for a small field $\vec{h} = (h_x, h_y, h_z)$, the effective low-energy Hamiltonian becomes
\begin{equation}
H = H_K - h_z \sum_{\mathbf{r} \in \mathbb{D}_z} i \tilde{b}_{\mathbf{r}}^z c_{\mathbf{r}}^{\phantom{\dag}}. \label{eq-H-3}
\end{equation}
Importantly, since the other two spin components have vanishing projections even at dangling sites $\mathbf{r} \in \mathbb{D}_z$, only the $z$ component of the field enters the low-energy Hamiltonian. We also note that the well-known ``chiral term'', which produces a bulk gap via second-neighbor matter-fermion hopping~\cite{Kitaev-2006}, only appears at third order in the field $\vec{h}$ and is thus negligible to the first-order coupling term in Eq.~(\ref{eq-H-3}).

\emph{Low-energy edge modes.}---To examine the low-energy edge dynamics, we first need to determine the relevant edge excitations of the effective Hamiltonian $H$ in Eq.~(\ref{eq-H-3}). Because of its quadratic nature, this Hamiltonian is readily diagonalized into the free-fermion form $H = \sum_n E_n^{\phantom{\dag}} f_n^{\dag} f_n^{\phantom{\dag}}$, where the fermion modes $f_n^{\dag}$ of excitation energies $E_n$ can be written as
\begin{equation}
f_n^{\dag} = \frac{1} {\sqrt{2}} \left[ \sum_{\mathbf{r}} \psi_{n, \mathbf{r}} \, c_{\mathbf{r}} + \sum_{\mathbf{r} \in \mathbb{D}_z} \phi_{n, \mathbf{r}}^{\phantom{\dag}} \, \tilde{b}_{\mathbf{r}}^z \right] \label{eq-f-1}
\end{equation}
in terms of the real-space coefficients $\psi_{n, \mathbf{r}}$ and $\phi_{n, \mathbf{r}}$ normalized by $\sum_{\mathbf{r}} |\psi_{n, \mathbf{r}}|^2 + \sum_{\mathbf{r} \in \mathbb{D}_z} |\phi_{n, \mathbf{r}}|^2 = 1$. These modes can be explicitly obtained as solutions of
\begin{equation}
\left[ H, f_n^{\dag} \right] = E_n^{\phantom{\dag}} f_n^{\dag}. \label{eq-f-2}
\end{equation}
Expressing $\mathbf{r} = a (X,Y)$ in terms of the bond length $a$ and the dimensionless coordinates $X$ and $Y$ being parallel and perpendicular to the zigzag edge, respectively, such that $Y = 0$ for all $\mathbf{r} \in \mathbb{D}_z$ and $Y > 0$ for all $\mathbf{r} \notin \mathbb{D}_z$ (see Fig.~\ref{fig-1}), we can search for edge-mode solutions with an edge momentum $k$ in the $X$ direction and a decay parameter $\lambda$ in the $Y$ direction,
\begin{align}
& \psi_{n, \mathbf{r} \in A} = \psi^A L^{-1/2} \lambda^{2(Y+1) / 3} e^{ikX / \sqrt{3}}, \nonumber \\
& \psi_{n, \mathbf{r} \in B} = -i \psi^B L^{-1/2} \lambda^{2Y / 3} e^{ikX / \sqrt{3}}, \label{eq-psi-1} \\
& \phi_{n, \mathbf{r} \in \mathbb{D}_z} = \phi^B L^{-1/2} e^{ikX / \sqrt{3}}, \nonumber
\end{align}
where $|k| < \pi$ and $|\lambda| < 1$, while $L$ is the number of dangling sites $\mathbf{r} \in \mathbb{D}_z$ (i.e., the length of the zigzag edge). The notation $\phi^B$ in Eq.~(\ref{eq-psi-1}) reflects that all dangling sites belong to the same sublattice $B$. From Eqs.~(\ref{eq-H-2})-(\ref{eq-psi-1}), we then readily obtain
\begin{subequations}
\begin{align}
& 2 K \left[ 1 + 2 \lambda^{-1} \cos (k/2) \right] \psi^B = E \psi^A, \label{eq-psi-2a} \\
& 2 K \left[ 1 + 2 \lambda \cos (k/2) \right] \psi^A = E \psi^B, \label{eq-psi-2b} \\
& 4 K \lambda \cos (k/2) \, \psi^A - 2 h_z \phi^B = E \psi^B, \label{eq-psi-2c} \\
& -2 h_z \psi^B = E \phi^B, \label{eq-psi-2d}
\end{align}
\end{subequations}
with $E \equiv E_n$. Note that the last two equations correspond to the dangling sites, thus containing $\phi^B$ and $h_z$.

\begin{figure}[t]
\includegraphics[width=0.9\columnwidth]{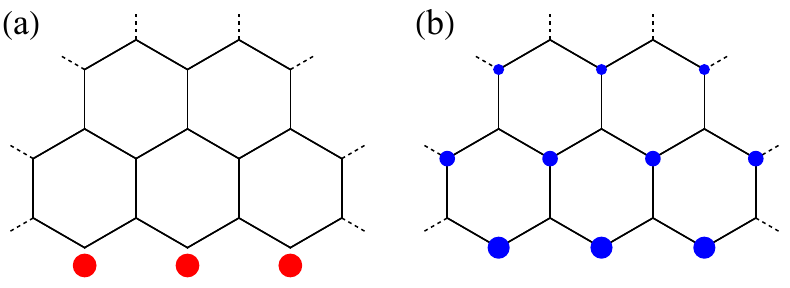}
\caption{Real-space wave functions of the zero-energy $h_z = 0$ edge modes $f_{b,k}^{\dag}$ (a) and $f_{c,k}^{\dag}$ (b). The area of each colored dot is proportional to the magnitude of the wave function at the given site, while red and blue colors indicate bond-fermion and matter-fermion character, respectively. Note that $f_{c,k}^{\dag}$ corresponds to $|\lambda_k| = 1/2$.} \label{fig-2}
\end{figure}

\begin{figure*}[t]
\includegraphics[width=1.8\columnwidth]{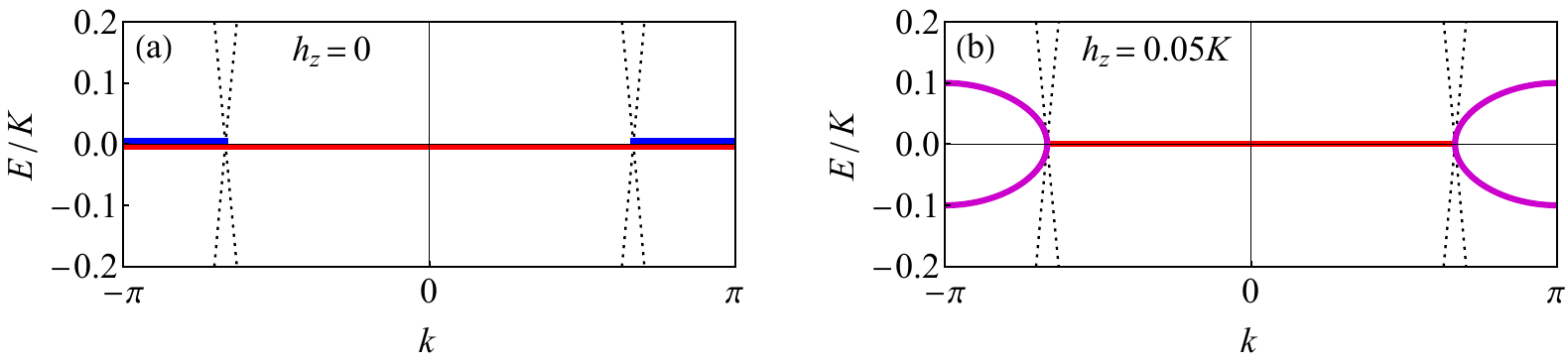}
\caption{Dispersions of the edge modes for $h_z = 0$ (a) and $h_z > 0$ (b). Solid lines with red, blue, and purple colors correspond to edge modes of bond-fermion, matter-fermion, and hybridized character, respectively. Bulk modes exist inside the wedges between the dotted lines.} \label{fig-3}
\end{figure*}

In the limit of no magnetic field, $h_z = 0$, the dangling bond fermions $\tilde{b}_{\mathbf{r}}^z$ are completely decoupled. Hence, we can readily identify a purely bond-fermion edge mode,
\begin{equation}
f_{b,k}^{\dag} = \frac{1} {\sqrt{2L}} \sum_{\mathbf{r} \in \mathbb{D}_z} e^{ikX / \sqrt{3}} \, \tilde{b}_{\mathbf{r}}^z, \label{eq-f-b}
\end{equation}
corresponding to $\phi_{b,k}^B = 1$ and $\psi_{b,k}^A = \psi_{b,k}^B = 0$, which is at exactly zero energy, $E = 0$, for all momenta $k$. By comparing Eqs.~(\ref{eq-psi-2b}) and (\ref{eq-psi-2c}), we next deduce $\psi^A = 0$, which implies that any matter-fermion edge mode must be confined to the $B$ sublattice. Using $\psi^A = 0$, we then find $E = 0$ from Eq.~(\ref{eq-psi-2b}) and $\lambda = -2 \cos(k/2)$ from Eq.~(\ref{eq-psi-2a}). Therefore, we can also identify a single normalized matter-fermion edge mode,
\begin{equation}
f_{c,k}^{\dag} = \sqrt{\frac{1 - \lambda_k^2} {2L}} \sum_{\mathbf{r} \in B} \lambda_k^{2Y / 3} e^{ikX / \sqrt{3}} \, c_{\mathbf{r}}, \label{eq-f-c}
\end{equation}
corresponding to $\psi_{c,k}^B = i \sqrt{1 - \lambda_k^2}$ and $\psi_{c,k}^A = \phi_{c,k}^B = 0$ with decay parameter $\lambda_k = -2 \cos(k/2)$. Due to the requirement $|\lambda_k| < 1$, this zero-energy edge mode only exists in the momentum range $2\pi / 3 < |k| < \pi$. Figure \ref{fig-2} shows the real-space wave functions of both zero-energy edge modes in Eqs.~(\ref{eq-f-b}) and (\ref{eq-f-c}). Note that, since $f_{b,-k}^{\dag} = f_{b,k}^{\phantom{\dag}}$ and $f_{c,-k}^{\dag} = f_{c,k}^{\phantom{\dag}}$, these edge modes are only physical for $k > 0$.

To understand the effect of a finite magnetic field, $h_z > 0$, it is instructive to consider the $h_z = 0$ edge modes in momentum space. As shown in Fig.~\ref{fig-3}(a), whereas the bond-fermion edge mode has local origin and extends to all edge momenta $k$, the matter-fermion edge mode is topological and connects the edge projections of the two bulk Dirac points~\cite{Perreault-2016,Mizoguchi-2019}. The coupling term $H_h = -h_z \sum_{\mathbf{r} \in \mathbb{D}_z} i \tilde{b}_{\mathbf{r}}^z c_{\mathbf{r}}^{\phantom{\dag}}$ in Eq.~(\ref{eq-H-3}) is then expected to hybridize the bond-fermion and the matter-fermion edge modes for momenta $2\pi / 3 < |k| < \pi$. By projecting this coupling term onto the fermion modes in Eqs.~(\ref{eq-f-b}) and (\ref{eq-f-c}), we find the reduced effective Hamiltonian
\begin{equation}
\hat{H}_h = 2 h_z \sum_{2\pi/3 < k < \pi} \sqrt{1 - \lambda_k^2} \left( i f_{c,k}^{\dag} f_{b,k}^{\phantom{\dag}} - i f_{b,k}^{\dag} f_{c,k}^{\phantom{\dag}} \right), \label{eq-H-4}
\end{equation}
and obtain hybridized modes $f_{\pm,k}^{\dag} = (f_{b,k}^{\dag} \pm i f_{c,k}^{\dag}) / \sqrt{2}$ with excitation energies $\pm E_k$ in terms of $E_k = 2 h_z \sqrt{1 - \lambda_k^2}$. The momentum dispersions of the hybridized edge modes are plotted in Fig.~\ref{fig-3}(b). We remark that the simple projection scheme above breaks down around $|k| = 2\pi/3$ due to hybridization with the bulk modes. However, our main results for the edge dynamics are not affected, as confirmed in the SM~\cite{SM} via a full solution of Eqs.~(\ref{eq-psi-2a})-(\ref{eq-psi-2d}) for $h_z > 0$.

\emph{Edge dynamics.}---We are interested in the low-energy spin dynamics at the edge of the Kitaev spin liquid. In particular, we consider the $z$ component of the single-site dynamical spin structure factor at a dangling site $\mathbf{r} \in \mathbb{D}_z$ without a $z$ bond:
\begin{equation}
S (\omega) \equiv S_{\mathbf{r} \in \mathbb{D}_z}^{zz} (\omega) = \frac{1} {2\pi} \int_{-\infty}^{\infty} dt \, e^{i \omega t} \, \langle \sigma_{\mathbf{r}}^z (t) \, \sigma_{\mathbf{r}}^z (0) \rangle. \label{eq-S-1}
\end{equation}
In contrast to other components~\cite{Baskaran-2007,Knolle-2014,Knolle-2015}, these dangling components do not involve flux creation and are expected to be finite even at energies below the flux gap $\Delta$~\cite{Kao-2021,Takahashi-2023,Kao-2024a,Kao-2024b}. In terms of the fermion modes in Eq.~(\ref{eq-f-1}), the inelastic (i.e., connected) part of the dangling spin structure factor in Eq.~(\ref{eq-S-1}) can be written in the Lehmann representation as
\begin{align}
S (\omega) &= \frac{1}{2} \sum_{m,n} \left| \langle \tilde{b}_{\mathbf{r}}^z c_{\mathbf{r}}^{\phantom{\dag}} f_m^{\dag} f_n^{\dag} \rangle \right|^2 \delta \left( \omega - E_m - E_n \right) \label{eq-S-2} \\
&= 2 \sum_{m,n} \left| \psi_{m, \mathbf{r}}^{\phantom{\dag}} \, \phi_{n, \mathbf{r}} - \psi_{n, \mathbf{r}}^{\phantom{\dag}} \, \phi_{m, \mathbf{r}} \right|^2 \delta \left( \omega - E_m - E_n \right). \nonumber
\end{align}
Note that the factor $1/2$ is included to avoid double counting when summing over the two-fermion states.

\begin{figure*}[t]
\includegraphics[width=2.1\columnwidth]{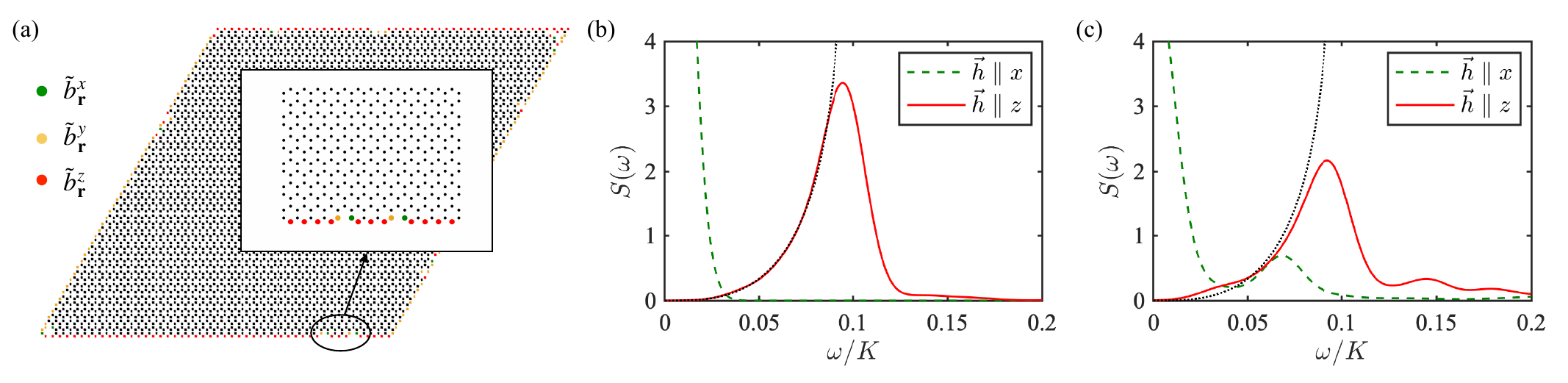}
\caption{(a) Finite honeycomb lattice with four disordered zigzag edges. (b-c) Low-energy spin structure factor $S (\omega)$ averaged over a single zigzag edge perpendicular to the $z$ bonds in the clean limit (b) and the disordered case (c) for a small magnetic field $h = 0.05K$ in the $z$ direction (red solid line) and the $x$ direction (green dashed line). The results in (c) correspond to disorder strength $P = 0.05$ and also include disorder averaging as described in the text. The black dotted line is the analytical result in Eq.~(\ref{eq-S-3}).} \label{fig-4}
\end{figure*}

Focusing on the $h_z > 0$ edge modes plotted in Fig.~\ref{fig-3}(b), the dangling spin structure factor in Eq.~(\ref{eq-S-2}) contains two distinct components: $S_{+,+} (\omega)$ and $S_{+,0} (\omega)$. The first component corresponds to two finite-energy excitations $f_{\pm,k}^{\dag}$ and $f_{\pm,k'}^{\dag}$. This component is featureless and has a small total spectral weight,
\begin{equation}
W_{+,+} = \frac{1} {2\pi^2} \int_{\frac{2\pi}{3}}^{\pi} dk \int_{\frac{2\pi}{3}}^{\pi} dk' \left| \psi_{c,k}^B \phi_{b,k'}^B - \psi_{c,k'}^B \phi_{b,k}^B \right|^2 < 0.01, \label{eq-W-1}
\end{equation}
since $\phi_{b,k}^B = 1$ and $\psi_{c,k}^B$ is a smooth function of $k$. In contrast, the second component $S_{+,0} (\omega)$ involves one finite-energy excitation $f_{\pm,k}^{\dag}$ and one zero-energy excitation $f_{b,k'}^{\dag}$. This component can be obtained analytically,
\begin{align}
S_{+,0} (\omega) &= \frac{1} {\pi^2} \int_{\frac{2\pi}{3}}^{\pi} dk \int_{0}^{\frac{2\pi}{3}} dk' \left| \psi_{c,k}^B \, \phi_{b,k'}^B \right|^2 \delta \left( \omega - E_k \right) \nonumber \\
&= \frac{\theta (\omega) \, \theta (2 h_z - \omega) \, \omega^3} {3\pi h_z^2 \sqrt{(12 h_z^2 + \omega^2) (4 h_z^2 - \omega^2)}}, \label{eq-S-3}
\end{align}
and has a significant total spectral weight,
\begin{equation}
W_{+,0} = \int_{0}^{2 h_z} d\omega \, S_{+,0} (\omega) = \frac{2} {\pi \sqrt{3}} - \frac{2}{9} \approx 0.15. \label{eq-W-2}
\end{equation}
Most importantly, as shown by Eq.~(\ref{eq-S-3}), this component exhibits a square-root divergence, $S_{+,0} (\omega) \propto (2 h_z - \omega)^{-1/2}$, at the top of its range, $\omega = 2 h_z$, due to the flat regions of the finite-energy edge modes around their extrema corresponding to $\pm E_{k=\pi} = \pm 2 h_z$ [see Fig.~\ref{fig-3}(b)]. Hence, the dynamical spin structure factor at the edge features a pronounced peak whose energy scales linearly with only the $z$ component of the magnetic field, reflecting the bond-directional spin interactions of the Kitaev honeycomb model.

\emph{Disorder effects.}---To confirm our analytical results for the edge dynamics and explore their robustness against edge disorder, we next present an exact numerical solution for a finite $50 \times 50$ system with an open boundary consisting of four disordered zigzag edges [see Fig.~\ref{fig-4}(a)]. The disordered boundary is generated by randomly removing each site along the boundary with probability $P$. For the resulting finite disordered system, there are dangling bond fermions $\tilde{b}_{\mathbf{r}}^x$ and $\tilde{b}_{\mathbf{r}}^y$ in addition to $\tilde{b}_{\mathbf{r}}^z$, and the low-energy Hamiltonian in Eq.~(\ref{eq-H-3}) generalizes to
\begin{equation}
H = H_K - \sum_{\alpha = x,y,z} h_{\alpha} \sum_{\mathbf{r} \in \mathbb{D}_{\alpha}} i \tilde{b}_{\mathbf{r}}^{\alpha} c_{\mathbf{r}}^{\phantom{\dag}}, \label{eq-H-5}
\end{equation}
where $\mathbf{r} \in \mathbb{D}_{\alpha}$ are dangling sites with no $\alpha$ bond connected to them. The dangling spin structure factors $S (\omega)$ can be computed by Eq.~(\ref{eq-S-2}) after generalizing $\tilde{b}_{\mathbf{r}}^z$ to $\tilde{b}_{\mathbf{r}}^{\alpha}$ and obtaining the fermion modes $f_n^{\dag}$ from Eq.~(\ref{eq-f-2}). To obtain a typical result, we average $S (\omega)$ over all dangling sites $\mathbf{r} \in \mathbb{D}_{\alpha}$ along a single zigzag edge perpendicular to the $z$ bonds, as well as over $150$ different random realizations of the edge disorder. We remark that the non-dangling spin structure factors involve flux creation and do not contribute at the lowest energies below the flux gap $\Delta \approx 0.26K$~\cite{Baskaran-2007,Knolle-2014,Knolle-2015}.

Our results for the low-energy spin structure factor are plotted in Fig.~\ref{fig-4}(b) for the clean limit ($P = 0$) and in Fig.~\ref{fig-4}(c) for the disordered case ($P > 0$). If the magnetic field is applied in the $z$ direction, $\vec{h} = (0,0,h)$, the spin structure factor exhibits a pronounced peak at $\omega = 2h$ in the clean limit, as predicted by Eq.~(\ref{eq-S-3}), which is broadened but still observable in the disordered case. For a magnetic field applied in the $x$ direction, $\vec{h} = (h,0,0)$, this finite-energy peak is instead replaced by a quasi-elastic response around $\omega = 0$, which is also consistent with Eq.~(\ref{eq-S-3}) in the limit of $h_z \to 0$. Therefore, we conclude that our key result, the pronounced peak at $\omega = 2h_z$, remains observable even in the presence of edge disorder.

\emph{Non-Kitaev interactions.}---Similar to a magnetic field, non-Kitaev perturbations introduce coupling between the dangling bond fermions $\tilde{b}_{\mathbf{r} \in \mathbb{D}_z}^z$ and the matter fermions $c_{\mathbf{r}}$. In contrast to a magnetic field, however, such time-reversal-symmetric perturbations necessarily couple two fermions on opposite honeycomb sublattices~\cite{You-2012,Song-2016}. Therefore, since the zero-energy $h_z = 0$ edge modes plotted in Fig.~\ref{fig-3}(a) are both confined to the $B$ sublattice (see Fig.~\ref{fig-2}), they remain at exactly zero energy for $h_z = 0$ and exhibit linear energy scaling for $h_z > 0$, even in the presence of non-Kitaev interactions~\cite{Footnote-1}. As an illustration of this general principle, the perturbed edge modes are explicitly obtained in the SM~\cite{SM} for a $\Gamma'$ perturbation that couples each dangling bond fermion to the matter fermions at the two neighboring sites. In turn, the linear energy scaling of the strong $S (\omega)$ peak, $\omega = \eta h_z$, is then expected to be robust against non-Kitaev interactions, although with a renormalized prefactor $\eta$~\cite{Zhang-2024}. This linear energy scaling reflects the projective sublattice and time-reversal symmetries of the Kitaev spin liquid~\cite{You-2012,Song-2016} and manifests along a zigzag edge for which all dangling sites belong to the same sublattice.

\emph{Discussion.}---We have proposed a sharp field-induced peak in the low-energy spin dynamics of the zigzag edge as a robust nanoscale signature of the Kitaev spin liquid. For a real monolayer candidate material like $\alpha$-RuCl$_3$, this peak may be experimentally observed by inelastic electron tunneling spectroscopy or color-center relaxometry.

In a tunneling setup~\cite{Yang-2023,Feldmeier-2020,Konig-2020,Bauer-2023,Takahashi-2023,Kao-2024a,Kao-2024b}, electrons tunnel between two metallic electrodes through a monolayer Kitaev spin liquid, and the sharp peak in the spin dynamics translates into a well-defined step in the tunneling conductance $dI/dV$ at bias voltage $V \sim g \mu_B B_z / e$ for a magnetic field $B_z$ in the appropriate direction~\cite{Footnote-2}. One-dimensional tunneling setups, such as the ``pincher gate'' in Ref.~\cite{Konig-2020}, are especially promising because they can closely follow the zigzag edge to maximize the edge signal while minimizing unwanted bulk contributions. If tunneling happens along a length $L$ of the edge, the step in the tunneling conductance can be estimated as~\cite{Bauer-2023,Kao-2024a}
\begin{equation}
\Delta \left( \frac{dI}{dV} \right) \sim \frac{L}{a} \times W_{+,0} \times \frac{\nu^2 t^4} {U^2} \times G_0, \label{eq-IV}
\end{equation}
where $G_0 = 2e^2 / h$ is the conductance quantum, $a$ is the lattice constant, $U$ is the charge gap of the spin liquid, $\nu$ is the density of states in the electrodes, and $t$ is the tunneling amplitude from each electrode to the spin liquid. For $a \sim 1$ nm, $t \sim 10$ meV, and $U \sim \nu^{-1} \sim 1$ eV, the step in $dI/dV$ is then comparable to $1$ nS~\cite{Yang-2023} for $L \gtrsim 10$ $\mu$m.

With color-center relaxometry~\cite{Rondin-2014,Casola-2018,Chatterjee-2019,Gottscholl-2021,Liu-2022}, spin fluctuations at a tunable intrinsic frequency are detected as an increase in the relaxation rate $1/T_1$. For a realistic disordered edge of the Kitaev spin liquid, the sharp peak at frequency $f \sim g \mu_B B_z / h$ spreads over a finite frequency range comparable to $f$ itself [see Fig.~\ref{fig-4}(c)]. Thus, if the color center is at distance $d$ from a monolayer edge, the increase in $1/T_1$ is approximately~\cite{Chatterjee-2019}
\begin{equation}
\Delta \left( \frac{1}{T_1} \right) \sim \frac{d}{a} \times \frac{W_{+,0}} {\mu_B B_z} \times \frac{1}{\hbar} \left( \frac{\mu_0 \mu_B^2} {d^3} \right)^2. \label{eq-T1}
\end{equation}
To produce an observable increase, $\Delta (1/T_1) \gtrsim 1$ ms$^{-1}$, for $B_z \lesssim 1$ T, the distance must be small enough, $d \lesssim 10$ nm, which is challenging for nitrogen-vacancy (NV) centers in diamond~\cite{Rondin-2014,Casola-2018}, but should be feasible with analogous color centers in hexagonal boron nitride (hBN)~\cite{Gottscholl-2021,Liu-2022}.

\emph{Acknowledgments.}---We are grateful to Benjamin Lawrie and Pramey Upadhyaya for enlightening discussions. This material is based upon work supported by the U.S. Department of Energy, Office of Science, National Quantum Information Science Research Centers, Quantum Science Center.



\clearpage

\begin{widetext}

\subsection{\large Supplemental Material}

\section{Fermion edge modes in the presence of a magnetic field}

Here we present a full solution of Eqs.~(7a)-(7d) in the main text to determine the exact fermion edge modes in the presence of a magnetic field $h_z$ and demonstrate that their extrema are still at energies $E = \pm 2h_z$. In the following, we consider zero-energy edge modes with $E = 0$ and finite-energy edge modes with $E \neq 0$ separately.

For any zero-energy edge modes, Eq.~(7d) readily implies $\psi^B = 0$, and Eq.~(7b) then requires $1 + 2 \lambda \cos (k/2) = 0$, thus giving a single edge mode with decay parameter $\lambda = -[2 \cos (k/2)]^{-1}$. Since the decay parameter must satisfy $|\lambda| < 1$, this edge mode only exists at edge momenta $|k| < 2\pi/3$. Furthermore, Eq.~(7c) gives $\psi^A = [2K \lambda \cos (k/2)]^{-1} h_z \phi^B = -(h_z / K) \, \phi^B$, which in turn shows that the edge mode is of predominantly bond-fermion character for $h_z \ll K$. This zero-energy edge mode precisely corresponds to the one shown in Fig.~3(b) of the main text.

For any finite-energy edge modes, the elimination of $\phi^B$ from Eqs.~(7c) and (7d) immediately leads to
\begin{equation}
4 K \lambda \cos (k/2) \, \psi^A = \left( E - \frac{4 h_z^2} {E} \right) \psi^B, \label{eq-S1}
\end{equation}
and the comparison of Eqs.~(7b) and (19) then implies $\psi^A = 2h_z^2 (K E)^{-1} \psi^B$. The substitution of this expression into Eq.~(7a) gives a single solution for the decay parameter,
\begin{equation}
\lambda = -\frac{2 K^2 \cos (k/2)} {K^2 - h_z^2}, \label{eq-S2}
\end{equation}
and the substitution of Eq.~(20) into the combination of Eqs.~(7a) and (7b) then gives two solutions for the energy,
\begin{equation}
E = \pm 2 K \sqrt{\left[ 1 + 2 \lambda^{-1} \cos (k/2) \right] \left[ 1 + 2 \lambda \cos (k/2) \right]} = \pm 2 h_z \sqrt{1 - \frac{4 K^2 \cos^2 (k/2)} {K^2 - h_z^2}}. \label{eq-S3}
\end{equation}
Because of the requirement $|\lambda| < 1$, these edge modes only exist at edge momenta $|k| > k_0 = 2 \arccos [(1 - h_z^2 / K^2) / 2]$. The absolute energies $|E|$ take their minimal values, $|E|_{\mathrm{min}} = 2h_z^2 / K$, at the critical momenta, $k = \pm k_0$, and their maximal values, $|E|_{\mathrm{max}} = 2h_z$, at the zone boundary, $k = \pm \pi$. Also, since $E \sim h_z$ according to Eq.~(21), we find $\psi^B \sim \phi^B$ from Eq.~(7d), which indicates that the edge modes have both bond-fermion and matter-fermion character. These finite-energy hybridized edge modes largely correspond to the ones shown in Fig.~3(b) of the main text. While the edge modes in the full solution connect to the bulk modes at critical momenta $|k| = k_0 > 2\pi/3$ and finite energies $|E|_{\mathrm{min}} = 2h_z^2 / K$, the simplified picture in the main text is recovered in the limit of $h_z \to 0$ because of $k_0 \to 2\pi/3$ and $|E|_{\mathrm{min}} \to 0$. In this limit, Eqs.~(20) and (21) also reduce to $\lambda = -2 \cos (k/2)$ and $E = \pm 2 h_z \sqrt{1 - \lambda^2}$, thereby reproducing the corresponding expressions in the main text. Even more importantly, the extremal energies of the edge modes correspond to $|E|_{\mathrm{max}} = 2h_z$ even for finite $h_z$, which implies that the pronounced peak in the spin structure factor still appears at energy $\omega = 2h_z$, as discussed in the main text.

\section{Fermion edge modes in the presence of a non-Kitaev perturbation}

Here we describe the fermion edge modes in the presence of a specific non-Kitaev perturbation---the $\Gamma'$ interaction:
\begin{equation}
\mathcal{H}_{\Gamma'} = \Gamma' \sum_{\alpha = x,y,z} \sum_{\beta \neq \alpha} \sum_{\langle \mathbf{r}, \mathbf{r}' \rangle_{\alpha}} \left( \sigma_{\mathbf{r}}^{\alpha} \sigma_{\mathbf{r}'}^{\beta} + \sigma_{\mathbf{r}}^{\beta} \sigma_{\mathbf{r}'}^{\alpha} \right). \label{eq-S4}
\end{equation}
This interaction has a nontrivial projection to the ground-state flux sector of the pure Kitaev model, which couples each dangling bond fermion $\tilde{b}_{\mathbf{r}}^z$ to the matter fermions $c_{\mathbf{r}}$ at the two neighboring sites. The effective low-energy Hamiltonian [see Eqs.~(2) and (3) in the main text] in the presence of the $\Gamma'$ interaction is then given by
\begin{equation}
H = K \sum_{\langle \mathbf{r} \in A, \mathbf{r}' \in B \rangle} i c_{\mathbf{r}} c_{\mathbf{r}'} + \Gamma' \sum_{\mathbf{r} \in \mathbb{D}_z} \left( i \tilde{b}_{\mathbf{r}}^z c_{\mathbf{r} + \mathbf{R}_x}^{\phantom{\dag}} - i \tilde{b}_{\mathbf{r}}^z c_{\mathbf{r} + \mathbf{R}_y}^{\phantom{\dag}} \right), \label{eq-S5}
\end{equation}
where the sites $\mathbf{r} + \mathbf{R}_x$ and $\mathbf{r} + \mathbf{R}_y$ are connected to the site $\mathbf{r}$ by $x$ and $y$ bonds, respectively. We search for edge-mode solutions of Eqs.~(4) and (5) in the main text using a generalized version of Eq.~(6) in the main text,
\begin{align}
& \psi_{n, \mathbf{r} \in A} = \psi^A L^{-1/2} \lambda^{2(Y+1) / 3} e^{ikX / \sqrt{3}}, \nonumber \\
& \psi_{n, \mathbf{r} \in B} = -i \left[ \psi^B + \bar{\psi}^B \delta_{Y,0} \right] L^{-1/2} \lambda^{2Y / 3} e^{ikX / \sqrt{3}}, \label{eq-S6} \\
& \phi_{n, \mathbf{r} \in \mathbb{D}_z} = \phi^B L^{-1/2} e^{ikX / \sqrt{3}}, \nonumber
\end{align}
and obtain a modified version of Eqs.~(7a)-(7d) in the main text,
\begin{subequations}
\begin{align}
& 2 K \left[ 1 + 2 \lambda^{-1} \cos (k/2) \right] \psi^B = E \psi^A, \label{eq-S7a} \\
& 2 K \left[ 1 + 2 \lambda \cos (k/2) \right] \psi^A = E \psi^B, \label{eq-S7b} \\
& 2 K \left[ 1 + 2 \lambda^{-1} \cos (k/2) \right] \psi^B + 4 K \lambda^{-1} \cos (k/2) \, \bar{\psi}^B - 4 \Gamma' \lambda^{-1} \sin (k/2) \, \phi^B = E \psi^A, \label{eq-S6c} \\
& 4 K \lambda \cos (k/2) \, \psi^A = E \left[ \psi^B + \bar{\psi}^B \right], \label{eq-S7d} \\
& -4 \Gamma' \lambda \sin (k/2) \, \psi^A = E \phi^B. \label{eq-S7e}
\end{align}
\end{subequations}
The comparison of Eqs.~(25a) and (25c) immediately gives
\begin{equation}
K \cos (k/2) \, \bar{\psi}^B - \Gamma' \sin (k/2) \, \phi^B = 0, \label{eq-S8}
\end{equation}
while the comparison of Eqs.~(25b) and (25d) readily implies
\begin{equation}
2 K \psi^A + E \bar{\psi}^B = 0. \label{eq-S9}
\end{equation}
From Eqs.~(25d), (25e), (26), and (27), a nontrivial solution for $(\psi^A, \psi^B, \bar{\psi}^B, \phi^B)$ requires
\begin{equation}
E^2 \left[ 2 \Gamma'^2 \lambda \sin^2 (k/2) - K^2 \cos (k/2) \right] = 0. \label{eq-S10}
\end{equation}
Hence, the decay parameter for any finite-energy ($E \neq 0$) solution becomes
\begin{equation}
\lambda = \frac{K^2 \cos (k/2)} {2 \Gamma'^2 \sin^2 (k/2)}, \label{eq-S11}
\end{equation}
and the substitution of Eq.~(29) into the combination of Eqs.~(25a) and (25b) implies $E^2 \geq 4 K^2$. In other words, there are no low-energy edge-mode solutions with absolute energies $0 < |E| \ll K$. For zero-energy ($E = 0$) solutions, Eq.~(28) is trivially satisfied, while Eq.~(27) immediately gives $\psi^A = 0$. From Eq.~(25a), we next obtain either $\lambda = -2 \cos (k/2)$ or $\psi^B = 0$, with Eq.~(26) providing a nontrivial solution for $(\bar{\psi}^B, \phi^B)$ in the latter case. Using these solutions, the generalizations of the fermion edge modes in Eqs.~(8) and (9) of the main text in the case of finite $\Gamma'$ are then given by
\begin{align}
& \bar{f}_{b,k}^{\,\dag} = \mathcal{N}_k^{(0)} \sum_{\mathbf{r} \in \mathbb{D}_z} e^{ikX / \sqrt{3}} \left[ K \cos (k/2) \, \tilde{b}_{\mathbf{r}}^z - i \Gamma' \sin (k/2) \, c_{\mathbf{r}}^{\phantom{\dag}} \right], \nonumber \\
& \bar{f}_{c,k}^{\,\dag} = \mathcal{N}_k^{(1)} \sum_{\mathbf{r} \in B} \lambda_k^{2Y / 3} e^{ikX / \sqrt{3}} \, c_{\mathbf{r}} + \mathcal{N}_k^{(2)} \sum_{\mathbf{r} \in \mathbb{D}_z} e^{ikX / \sqrt{3}} \left[ K \cos (k/2) \, \tilde{b}_{\mathbf{r}}^z - i \Gamma' \sin (k/2) \, c_{\mathbf{r}}^{\phantom{\dag}} \right], \label{eq-S12}
\end{align}
where the decay parameter is $\lambda_k = -2 \cos (k/2)$, and the overall coefficients $\mathcal{N}_k^{(0,1,2)}$ are chosen so that the two fermion modes are normalized and orthogonal to each other. Note that the two respective fermion modes are no longer of purely bond-fermion and purely matter-fermion character. Importantly, however, since both fermion modes are at exactly zero energy in the absence of a magnetic field, projecting a small magnetic-field perturbation onto these fermion modes gives excitation energies that scale linearly with the field strength, as described in the main text.

\clearpage

\end{widetext}



\begin{references}

\bibitem{Kitaev-2006} A. Y. Kitaev, Ann. Phys. \textbf{321}, 2 (2006).
\bibitem{Kitaev-2003} A. Y. Kitaev, Ann. Phys. \textbf{303}, 2 (2003).
\bibitem{Nayak-2008} C. Nayak, S. H. Simon, A. Stern, M. Freedman, and S. Das Sarma, Rev. Mod. Phys. \textbf{80}, 1083 (2008).
\bibitem{Plumb-2014} K. W. Plumb, J. P. Clancy, L. J. Sandilands, V. V. Shankar, Y. F. Hu, K. S. Burch, H.-Y. Kee, and Y.-J. Kim, Phys. Rev. B \textbf{90}, 041112 (2014).
\bibitem{Kubota-2015} Y. Kubota, H. Tanaka, T. Ono, Y. Narumi, and K. Kindo, Phys. Rev. B \textbf{91}, 094422 (2015).
\bibitem{Sandilands-2015} L. J. Sandilands, Y. Tian, K. W. Plumb, Y.-J. Kim, and K. S. Burch, Phys. Rev. Lett. \textbf{114}, 147201 (2015).
\bibitem{Sears-2015} J. A. Sears, M. Songvilay, K. W. Plumb, J. P. Clancy, Y. Qiu, Y. Zhao, D. Parshall, and Y.-J. Kim, Phys. Rev. B \textbf{91}, 144420 (2015).
\bibitem{Majumder-2015} M. Majumder, M. Schmidt, H. Rosner, A. A. Tsirlin, H. Yasuoka, and M. Baenitz, Phys. Rev. B \textbf{91}, 180401(R) (2015).
\bibitem{Johnson-2015} R. D. Johnson, S. C. Williams, A. A. Haghighirad, J. Singleton, V. Zapf, P. Manuel, I. I. Mazin, Y. Li, H. O. Jeschke, R. Valent\'i, and R. Coldea, Phys. Rev. B \textbf{92}, 235119 (2015).
\bibitem{Sandilands-2016} L. J. Sandilands, Y. Tian, A. A. Reijnders, H.-S. Kim, K. W. Plumb, Y.-J. Kim, H.-Y. Kee, and K. S. Burch, Phys. Rev. B \textbf{93}, 075144 (2016).
\bibitem{Banerjee-2016} A. Banerjee, C. A. Bridges, J.-Q. Yan, A. A. Aczel, L. Li, M. B. Stone, G. E. Granroth, M. D. Lumsden, Y. Yiu, J. Knolle, S. Bhattacharjee, D. L. Kovrizhin, R. Moessner, D. A. Tennant, D. G. Mandrus, and S. E. Nagler, Nat. Mater. \textbf{15}, 733 (2016).
\bibitem{Banerjee-2017} A. Banerjee, J. Yan, J. Knolle, C. A. Bridges, M. B. Stone, M. D. Lumsden, D. G. Mandrus, D. A. Tennant, R. Moessner, and S. E. Nagler, Science \textbf{356}, 1055 (2017).
\bibitem{Do-2017} S.-H. Do, S.-Y. Park, J. Yoshitake, J. Nasu, Y. Motome, Y. S. Kwon, D. T. Adroja, D. J. Voneshen, K. Kim, T.-H. Jang, J.-H. Park, K.-Y. Choi, and S. Ji, Nat. Phys. \textbf{13}, 1079 (2017).
\bibitem{Banerjee-2018} A. Banerjee, P. Lampen-Kelley, J. Knolle, C. Balz, A. A. Aczel, B. Winn, Y. Liu, D. Pajerowski, J. Yan, C. A. Bridges, A. T. Savici, B. C. Chakoumakos, M. D. Lumsden, D. A. Tennant, R. Moessner, D. G. Mandrus, and S. E. Nagler, npj Quantum Materials \textbf{3}, 8 (2018).
\bibitem{Kasahara-2018} Y. Kasahara, T. Ohnishi, Y. Mizukami, O. Tanaka, S. Ma, K. Sugii, N. Kurita, H. Tanaka, J. Nasu, Y. Motome, T. Shibauchi, and Y. Matsuda, Nature \textbf{559}, 227 (2018).
\bibitem{Balz-2019} C. Balz, P. Lampen-Kelley, A. Banerjee, J. Yan, Z. Lu, X. Hu, S. M. Yadav, Y. Takano, Y. Liu, D. A. Tennant, M. D. Lumsden, D. Mandrus, and S. E. Nagler, Phys. Rev. B \textbf{100}, 060405(R) (2019).
\bibitem{Yokoi-2021} T. Yokoi, S. Ma, Y. Kasahara, S. Kasahara, T. Shibauchi, N. Kurita, H. Tanaka, J. Nasu, Y. Motome, C. Hickey, S. Trebst, and Y. Matsuda, Science \textbf{373}, 568 (2021).
\bibitem{Bruin-2022} J. A. N. Bruin, R. R. Claus, Y. Matsumoto, N. Kurita, H. Tanaka, and H. Takagi, Nat. Phys. \textbf{18}, 401 (2022).
\bibitem{Imamura-2024} K. Imamura, S. Suetsugu, Y. Mizukami, Y. Yoshida, K. Hashimoto, K. Ohtsuka, Y. Kasahara, N. Kurita, H. Tanaka, P. Noh, J. Nasu, E.-G. Moon, Y. Matsuda, and T. Shibauchi, Sci. Adv. \textbf{10}, eadk3539 (2024).
\bibitem{Yamashita-2020} M. Yamashita, J. Gouchi, Y. Uwatoko, N. Kurita, and H. Tanaka, Phys. Rev. B \textbf{102}, 220404(R) (2020).
\bibitem{Czajka-2021} P. Czajka, T. Gao, M. Hirschberger, P. Lampen-Kelley, A. Banerjee, J. Yan, D. G. Mandrus, S. E. Nagler, and N. P. Ong, Nat. Phys. \textbf{17}, 915 (2021).
\bibitem{Lefrancois-2022} \'E. Lefran\c{c}ois, G. Grissonnanche, J. Baglo, P. Lampen-Kelley, J.-Q. Yan, C. Balz, D. Mandrus, S. E. Nagler, S. Kim, Y.-J. Kim, N. Doiron-Leyraud, and L. Taillefer, Phys. Rev. X \textbf{12}, 021025 (2022).
\bibitem{Czajka-2023} P. Czajka, T. Gao, M. Hirschberger, P. Lampen-Kelley, A. Banerjee, N. Quirk, D. G. Mandrus, S. E. Nagler, and N. P. Ong, Nat. Mater. \textbf{22}, 36 (2023).
\bibitem{Yang-2023} B. Yang, Y. M. Goh, S. H. Sung, G. Ye, S. Biswas, D. A. S. Kaib, R. Dhakal, S. Yan, C. Li, S. Jiang, F. Chen, H. Lei, R. He, R. Valent\'i, S. M.Winter, R. Hovden, and A.W. Tsen, Nat. Mater. \textbf{22}, 50 (2023).
\bibitem{Chaves-2020} A. Chaves, J. G. Azadani, H. Alsalman, D. R. da Costa, R. Frisenda, A. J. Chaves, S. H. Song, Y. D. Kim, D. He, J. Zhou, A. Castellanos-Gomez, F. M. Peeters, Z. Liu, C. L. Hinkle, S.-H. Oh, P. D. Ye, S. J. Koester, Y. H. Lee, P. Avouris, X. Wang, and T. Low, npj 2D Materials and Applications \textbf{4}, 29 (2020).
\bibitem{You-2012} Y.-Z. You, I. Kimchi, and A. Vishwanath, Phys. Rev. B \textbf{86}, 085145 (2012).
\bibitem{Song-2016} X.-Y. Song, Y.-Z. You, and L. Balents, Phys. Rev. Lett. \textbf{117}, 037209 (2016).
\bibitem{Zhang-2024} S.-S. Zhang, G. B. Hal\'asz, and C. D. Batista, in preparation.
\bibitem{Feldmeier-2020} J. Feldmeier, W. Natori, M. Knap, and J. Knolle, Phys. Rev. B \textbf{102}, 134423 (2020).
\bibitem{Konig-2020} E. J. K\"onig, M. T. Randeria, and B. J\"ack, Phys. Rev. Lett. \textbf{125}, 267206 (2020).
\bibitem{Bauer-2023} T. Bauer, L. R. D. Freitas, R. G. Pereira, and R. Egger, Phys. Rev. B \textbf{107}, 054432 (2023).
\bibitem{Takahashi-2023} M. O. Takahashi, M. G. Yamada, M. Udagawa, T. Mizushima, and S. Fujimoto, Phys. Rev. Lett. \textbf{131}, 236701 (2023).
\bibitem{Kao-2024a} W.-H. Kao, N. B. Perkins, and G. B. Hal\'asz, Phys. Rev. Lett. \textbf{132}, 136503 (2024).
\bibitem{Kao-2024b} W.-H. Kao, G. B. Hal\'asz, and N. B. Perkins, Phys. Rev. B \textbf{109}, 125150 (2024).
\bibitem{Rondin-2014} L. Rondin, J.-P. Tetienne, T. Hingant, J.-F. Roch, P. Maletinsky, and V. Jacques, Rep. Prog. Phys. \textbf{77}, 056503 (2014).
\bibitem{Casola-2018} F. Casola, T. van der Sar, and A. Yacoby, Nat. Rev. Mater. \textbf{3}, 17088 (2018).
\bibitem{Chatterjee-2019} S. Chatterjee, J. F. Rodriguez-Nieva, and E. Demler, Phys. Rev. B \textbf{99}, 104425 (2019).
\bibitem{Gottscholl-2021} A. Gottscholl, M. Diez, V. Soltamov, C. Kasper, D. Krau{\ss}e, A. Sperlich, M. Kianinia, C. Bradac, I. Aharonovich, and V. Dyakonov, Nat. Commun. \textbf{12}, 4480 (2021).
\bibitem{Liu-2022} W. Liu, N.-J. Guo, S. Yu, Y. Meng, Z.-P. Li, Y.-Z. Yang, Z.-A. Wang, X.-D. Zeng, L.-K. Xie, and Q. Li, Mater. Quantum. Technol. \textbf{2}, 032002 (2022).
\bibitem{Kao-2021} W.-H. Kao, J. Knolle, G. B. Hal\'asz, R. Moessner, and N. B. Perkins, Phys. Rev. X \textbf{11}, 011034 (2021).
\bibitem{Perreault-2016} B. Perreault, J. Knolle, N. B. Perkins, and F. J. Burnell, Phys. Rev. B \textbf{94}, 104427 (2016).
\bibitem{Mizoguchi-2019} T. Mizoguchi and T. Koma, Phys. Rev. B \textbf{99}, 184418 (2019).
\bibitem{SM} See the Supplemental Material for full analytical solutions of the fermion edge modes in the presence of a magnetic field and a non-Kitaev perturbation.
\bibitem{Baskaran-2007} G. Baskaran, S. Mandal, and R. Shankar, Phys. Rev. Lett. \textbf{98}, 247201 (2007).
\bibitem{Knolle-2014} J. Knolle, D. L. Kovrizhin, J. T. Chalker, and R. Moessner, Phys. Rev. Lett. \textbf{112}, 207203 (2014).
\bibitem{Knolle-2015} J. Knolle, D. L. Kovrizhin, J. T. Chalker, and R. Moessner, Phys. Rev. B \textbf{92}, 115127 (2015).
\bibitem{Footnote-1} While four-fermion terms of strength $J$ can give a finite energy shift and thus modify the linear scaling for $h_z < J$, such four-fermion terms are only generated at higher orders of the dominant non-Kitaev interactions~\cite{Winter-2017,Maksimov-2020} and are hence expected to be very small.
\bibitem{Footnote-2} Note that the factor $\eta \sim 2$ from $\omega = \eta h_z$ approximately cancels with the spin factor $S = 1/2$.
\bibitem{Winter-2017} S. M. Winter, A. A. Tsirlin, M. Daghofer, J. van den Brink, Y. Singh, P. Gegenwart, and R. Valent\'i, J. Phys. Condens. Matter \textbf{29}, 493002 (2017).
\bibitem{Maksimov-2020} P. A. Maksimov and A. L. Chernyshev, Phys. Rev. Research \textbf{2}, 033011 (2020).

\end{references}
\end{document}